\documentclass[final,5p,times,twocolumn,authoryear]{elsarticle}
\usepackage{lineno}
\usepackage[utf8]{inputenc}
\usepackage{soul}
\usepackage{graphics,graphicx,subcaption}
\usepackage{amsfonts,latexsym,cancel}
\usepackage{ulem}
\usepackage{xcolor}
\usepackage{amssymb}
\usepackage{lipsum}
\usepackage{lineno}
 \usepackage{amsmath}
 \usepackage{multirow}
\usepackage{graphicx}
\usepackage{txfonts}
\usepackage{natbib}
 \usepackage{rotating}
\usepackage[bookmarks=false]{hyperref}

\usepackage{amssymb}
\usepackage{amsmath}

\journal{High Energy Astrophysics}

\begin{document}

\begin{frontmatter}



\title{Exploring Temperature Influences on Gravitational Wave Production in Binary White Dwarfs}


\author[a]{S\'ilvia P. Nunes}\ead{nunes.silvia@ce.uerj.br}
\author[label2,label3]{Jos\'e D. V. Arba\~nil}
\author[label4]{C\'esar H. Lenzi}
\author[label5,label6]{Jaziel G. Coelho}


\affiliation[a]{organization={Instituto de Física Armando Dias Tavares, Universidade do Estado do Rio de Janeiro, Rua São Francisco Xavier 524, 20550-900 Rio de Janeiro, RJ, Brazil}}
\affiliation[label2]{organization={Departamento de Ciencias, Universidad Privada del Norte, Avenida el Sol 461 San Juan de Lurigancho, 15434 Lima,  Peru}}
\affiliation[label3]{organization={Facultad de Ciencias Físicas, Universidad Nacional Mayor de San Marcos, Avenida Venezuela s/n Cercado de Lima, 15081 Lima,  Peru}}
\affiliation[label4]{organization={Departamento de Física, Instituto Tecnológico de Aeronáutica, Centro Técnico Aeroespacial, 12228-900 São José dos Campos, São Paulo, Brazil}}
\affiliation[label5]{organization={ Divisão de Astrofísica, Instituto Nacional de Pesquisas Espaciais, Avenida dos Astronautas 1758, São José dos Campos, SP 12227-010, Brazil}}
\affiliation[label6]{organization={Núcleo de Astrofísica e Cosmologia (Cosmo-Ufes) \& Departamento de Física, Universidade Federal do Espírito Santo, 29075–910 Vitória, ES, Brazil}}

\begin{abstract}
This study investigates the conditions under which gravitational waves (GWs) are emitted during the merger of hot white dwarfs (WDs) in a binary system. Traditionally, these systems consist of two low-mass stars or a more massive WD paired with a less massive companion.  In addition, recent work has investigated the possibility that double white dwarf (DWD) mergers are possibly the leading formation channel of massive,
rapidly rotating, high-field magnetic WDs , particularly SDSS J221141.80 + 113604.4 (hereafter J2211+1136) and ZTF J190132.9 + 145808.7 (hereafter J1901 + 14588). Motivated by these findings and the Laser Interferometer Space Antenna (LISA) prospects, this study aims to calculate the tidal Love number, the dimensionless tidal deformability, as well as the frequency and amplitude of GWs of hot WDs. The results indicate that  the tidal deformability is more pronounced in stars with higher central temperatures and lower masses,  which would lead to reduced emission of GWs. In contrast, more massive stars exhibit less deformability, making them prime candidates for generating stronger GWs. Additionally, the analysis of frequency and amplitude reveals that the frequencies of high-mass binaries are smaller and evolve more rapidly, reaching a limit that aligns with the operational detection capabilities of LISA during its initial phase. 

\end{abstract}




\end{frontmatter}




\section{Introduction} \label{sec:intro}

Gravitational waves (GWs), predicted by Albert Einstein in his General Theory of Relativity, represent ripples in the fabric of spacetime created by extreme astrophysical phenomena, including the mergers of compact objects such as black holes \citep{Abbott_2016, Abbott_2016b}, neutron stars (NSs) \citep{Abbott_2017c, Abbott_2017b, Abbott_2023}, and white dwarfs (WDs) \citep{Kilic_2011, Perot_2022}. The Laser Interferometer Gravitational Wave Observatory (LIGO) was the first facility to successfully detect these waves, focusing primarily on more compact objects. LIGO operates effectively across frequencies from $10\,\text{Hz}$ to $100\,\text{Hz}$ \citep{LIGO_2015}, which adequately captures events involving black holes and neutron stars. However, WDs, being less compact, generate GW with amplitude and frequency lower than those observed in the LIGO sensitivity curve.

Although LIGO has limitations in detecting GWs from WDs, the Laser Interferometer Space Antenna (LISA) is anticipated to address this gap \citep{LISA_NASA}. Scheduled for launch in $2030$ s, LISA is designed to detect extremely faint signals, requiring a highly sensitive instrument. Specifically, LISA aims to measure relative position shifts smaller than the diameter of a helium nucleus over a distance of one million miles, achieving a strain sensitivity of $1\times10^{-20}$ at frequencies around one millihertz, as reported by NASA. During its initial four years of operation, LISA is expected to capture GWs that are not detectable by LIGO, thus providing valuable information to the dynamics of WD mergers and other low-frequency astrophysical events \citep{Wolz_2020}.

Being WDs an important target for LISA operation, recent studies have focused on proposing prospects for founding.  In $2005$, \cite{LorenAguilar_2005} proposed the calculation of GWs from binary systems composed of two WDs. These waves are generated during the merging process between the stars. Their study used a smoothed particle hydrodynamics (SPH) code to model the temporal evolution of coalescing WDs, revealing that the most notable feature of the gravitational wave signal is its sudden disappearance after the merger ~\citep[see][for details]{LorenAguilar_2005}. The ``chirping" phase before the merger will be detectable by instruments such as the LISA, making these binaries identifiable.

\cite{Perot_2022} says that GW in space using three spacecraft detectors can enhance our understanding of the interior of WDs in binary systems by revealing tidal effects in GW signals, concluding that some observed binaries may have WDs that are partially crystallized. Their research indicates that the crystallized core's elasticity significantly reduces tidal deformability, mostly in low-mass stars. \cite{Gergousi_2022} developed an analysis framework to extract information about the properties of the Milky Way by characterizing the spectral shape or the residual foreground signal of DWD measured by LISA. Their approach involved simulating catalogs of sources based on a fiducial binary population synthesis model, which included varying parameters like total stellar mass and the shape of the Galaxy. They estimated the residual foreground signal after accounting for the loudest sources, using methodologies from prior studies.  

In this work, we investigate the conditions for GW emission during the merger of WDs, inspired by existing research on neutron star mergers and the prospects of detection with LISA. Traditionally, binary systems consist of a more massive WD paired with a less massive companion \citep{Brown_2013,Brown_2016,Gianninas_2014}. However, recent studies by \cite{Sousa_2022} have concluded that the observed parameters of J2211+1136 and J1901+1458 are consistent with a DWD merger origin. Interestingly, the derived parameters of the merging DWDs are in line with those of known DWDs, such as NLTT 12758~\citep[see][for details]{2017MNRAS.466.1127K}, which further supports the connection between massive, rapidly rotating, highly magnetic WDs and DWD mergers.
The potential for merging massive WDs has been anticipated since the work of \cite{Benz_1990}, and further studies, including \cite{Kilic_2013}, suggest that these mergers could serve as progenitors for Type Ia supernovae. Also, DWD mergers are significant astrophysical events expected to produce massive, highly magnetized white dwarfs (WDs). Despite predictions that they should be as numerous in the sky as other transient sources, their detection has remained elusive, even for the most advanced transient surveys. \citet{Sousa_2023} anticipates that the Legacy Survey of Space and Time (LSST) at the Vera C. Rubin Observatory will have the capability to detect this new class of astrophysical sources. The study forecasts that emissions associated with WD binary mergers, spanning wavelengths from infrared to ultraviolet, will be observed at an exceptionally high rate, potentially up to a thousand events per year~\citep[see][for details]{Sousa_2023}. To model the stellar structure configurations of hot WDs, we use the equation of state (EoS) with temperature effects from \cite{Nunes_2021} and solve the static equilibrium configurations \citep{Tolman_1939,OV_1939}. Our research computes parameters such as the tidal Love number and dimensionless tidal deformability for individual WDs as well as the frequency $f$ and amplitude $\mathcal{A}$ of GWs produced by two WDs in a binary system during the merger process, exploring the influence of temperature and varying chirp masses.

The following sections are divided as follows: in Sec.~\ref{eos} the equation of state which describes the fluid contained and the temperature profile in white dwarf are presented. In Sec. \ref{general_relativistic_equation}, the static equilibrium equation and the tidal deformability equations are shown. In Sec.~\ref{numerical_results} the tidal Love number and dimensionless tidal deformability for individual WDs, as well as the frequency $f$ and amplitude $\mathcal{A}$ of gravitational waves coming from two WDs in a binary system during the merger process are explored. In Sec.~\ref{conclusion_ofthework} we conclude. Finally, throughout the article, we employ the metric signature $(-,+,+,+)$ and we use the geometrized units $c=1=G$.

\section{The equation of state}\label{eos}

To describe the fluid contained in the stellar structure, we adopt the EoS for the interstellar fluid as proposed by \cite{Timmes_1999} and extended by \cite{Nunes_2021}. Since we consider high temperatures where the fluid is in a molten state, the lattice contribution typically present at lower temperatures is neglected. The chosen EoS  account for contributions from nucleons, electrons, and radiation, which are essential components of the interstellar fluid. The nucleons, represented by protons and neutrons, contribute to the fluid's overall mass and energy density. Electrons, although being partially degenerated in hot environments, significantly influence the fluid's pressure. In addition, radiation, typically in the form of photons, plays an important role in pressure and energy at stellar edge. By incorporating these constituents into EoS, we obtain a comprehensive representation of the thermodynamic properties of the interstellar fluid, enabling us to model the intricate structure and behavior of stars under the considered high-temperature conditions,
\begin{eqnarray}\label{ch_pressure}
P&=&P_n+P_e+P_\gamma,\\\label{ch_e}
\varepsilon&=&\varepsilon_n+\varepsilon_e+\varepsilon_\gamma,
\end{eqnarray}
where the subindex $n$, $e$, and $\gamma$ represent nucleons, electrons and photons contributions, respectively. Regarding the white dwarfs' temperature, the stellar interior is an isothermal core with constant temperature and a non-degenerate envelope, which has a temperature distribution due to the luminosity. In this work, we follow the temperature-density relation used in \citep{Nunes_2021}, i.e.,
\begin{equation}
    T\propto\rho^{2/3},
\end{equation}
being this relationship valid only for densities below the degeneracy threshold.

\section{General relativistic equations}\label{general_relativistic_equation}

\subsection{Static equilibrium equation}\label{gr}

To describe static spherically symmetric WDs, we employ the following line element,
\begin{equation}\label{metric}
ds^2=-e^{\nu}dt^2+e^{\lambda}dr^2+r^2d\theta^2+r^2\sin^2\theta d\phi^2,
\end{equation}
with $t, r, \theta,$ and $\phi$ being the Schwarzschild-like coordinates. The metric functions $\nu=\nu(r)$, and $\lambda=\lambda(r)$ depend on the radial coordinate $r$ alone. 

By using the non-null Einstein field equations, the stellar equilibrium configuration equations [also called \cite{Tolman_1939,OV_1939} equations] are derived. These equations can be written in the form,
\begin{eqnarray}
\hspace{-0.7cm}\frac{dm}{dr}&=&4\pi\varepsilon r^2,\label{dm}\\
\hspace{-0.7cm}\frac{dP}{dr}&=&-(P+\varepsilon)\left[4\pi rP+\frac{m}{r^2}\right]e^\lambda,\label{dp}\\
\end{eqnarray}
with the metric function $e^\lambda$ of the form,
\begin{equation}\label{eq_lambda}
e^{\lambda}=\left[1-\frac{2m}{r}\right]^{-1}.
\end{equation}
The function $m=m(r)$ represents the mass enclosed within the sphere radius $r$.

The equilibrium configurations are obtained integrating the differential equations \eqref{dm}-\eqref{dp} from the center ($r=0$) of the star -- where $m(r=0)=0$, 
$\rho(r=0)=\rho_c$, and $p(r=0)=p_c$-- until the star's surface ($r=R$) -- which is attained when $p(r=R)=0$. Moreover, in $r=R$, since the interior metric matches smoothly with the exterior Schwarzschild metric, we have:
\begin{equation}
    e^{\nu(R)}=e^{-\lambda(R)}=1-\frac{2M}{R},
\end{equation}
where $M$ stands for the total stellar mass.

\subsection{Tidal Love number and deformability}

{ The study of binary compact stars often involves the analysis of the tidal love number and deformability parameter. These measurements quantify the perturbation of the quadrupole moment in one of the stars of the binary system due to the external field created by its companion. These parameters correlate as follows:}

\begin{equation}
    \lambda = k_2 \frac{2}{3}\lambda R^{5},
\end{equation}
where the quadripolar Love number $k_2$ can be calculated through the use of equality,
\begin{eqnarray}\label{k2_parameter}
k_2=\frac{8C^5}{5}(1-2C)^2\left[2+C(y_R-1)-y_R\right]\times\nonumber \\\{2C\left[6-3y_R+ 3C(5y_R-8)\right] \nonumber \\ +4C^3\left[13-11y_R+C(3y_R-2)\right.
\left.+2C^2(1+y_R)\right]\nonumber \\ +3(1-2C^2)\times[2-y_R+2(y_R-1)]\ln(1-2C)\}^{-1},
\end{eqnarray}
where $y_R=y(r=R)$ is a function that satisfies the first-order differential equation,
\begin{equation}\label{y_equation}
y'r+y^2+y(C_0r-1)+C_1r^2=0,
\end{equation}
with
\begin{eqnarray}
C_0=\frac{2m}{r^2}e^{\lambda}+4\pi e^{\lambda}(p-\rho)r+\frac{2}{r},\label{C0}\\
C_1=4\pi e^{\lambda}\left[5\rho+9p+\frac{\rho+p}{c_s^2}\right]-\frac{6}{r^2}e^{\lambda}-\nu'^2,\label{C1}
\end{eqnarray}
where $c_s$ represents the speed of sound and is calculated by the relation $c_s^2=\frac{dp}{d\rho}$.

Additionally, by using the Love number parameter $k_2$, the dimensionless tidal deformability $\Lambda$ can be determined through the following relationship,
\begin{equation}\label{tidal_lambda}
    \Lambda=\frac{2k_2}{3C^5},
\end{equation}
with $C=M/R$ standing the compactness parameter; review, e.g., the articles of \cite{hinderer_2008}, \cite{damour2009} and \cite{hinderer_2010}.

As reported by \cite{Postnikov_2010}, in the Newtonian limit -where the compact parameter is small $C\rightarrow0$, $p<<\rho$, and $\rho r^2<<1$- the tidal Love number equation (\ref{k2_parameter}) and Riccati differential equation (\ref{y_equation}) can be placed respectively into the form,
\begin{eqnarray}
    k_2=\frac{1}{2}\left(\frac{2-y_R}{3+y_R}\right),\label{k2_parameter_reduced}\\
    ry'+y^2+y-6+4\pi r^2\frac{\rho}{c_s^2}=0,\label{y_equation_reduced}
\end{eqnarray}
where, in the star's center ($r=0$), the function $y$ takes the value $y(0)=2$. In this way, equations \eqref{tidal_lambda} and \eqref{k2_parameter_reduced} will allow us to analyze the dimensionless tidal deformability and the tidal Love number of WDs with temperature, respectively.

\section{Results}\label{numerical_results}

\subsection{Numerical method scheme}

 Once the EoS and temperature profile is defined, the stellar equilibrium equations \eqref{dm}-\eqref{eq_lambda} and the tidal deformability equation \eqref{y_equation_reduced} are numerically solved from the center $(r=0)$ towards the star's surface $(r=R)$ employing the fourth-order Runge-Kutta method to investigate the influence of the central temperature on some physical parameters of WD. Namely, after the numerical integration, in the following subsection, the results presented in the form of figures show the influence of central temperature $T_c$ on the tidal love number $k_2$ and the dimensionless tidal deformability $\Lambda$ of WDs and on $\Lambda_1\times\Lambda_2$, and ${\tilde\Lambda}$ in a binary WD system.

\subsection{Effect of temperature on tidal Love number and deformability of white dwarfs}\label{Results}

 The tidal Love number $k_2$ and tidal deformability provide valuable insights into the stellar internal structure and elasticity, thus revealing the star's response to external gravitational forces. In binary star systems, the tidal Love number helps understand the orbital evolution and the gravitational wave signals emitted \citep{Peters_1964,Huang_2020,Perot_2022}. In close binary systems, tidal interactions can cause a shrink in the orbit over time, leading to orbital decay. The tidal Love number influences this decay rate, determining how efficiently the tidal energy is dissipated within the star (see \citep{Damour_2009,Binnington_2009}).

\begin{figure}[!ht] 
\begin{center}
\includegraphics[width=1\linewidth]{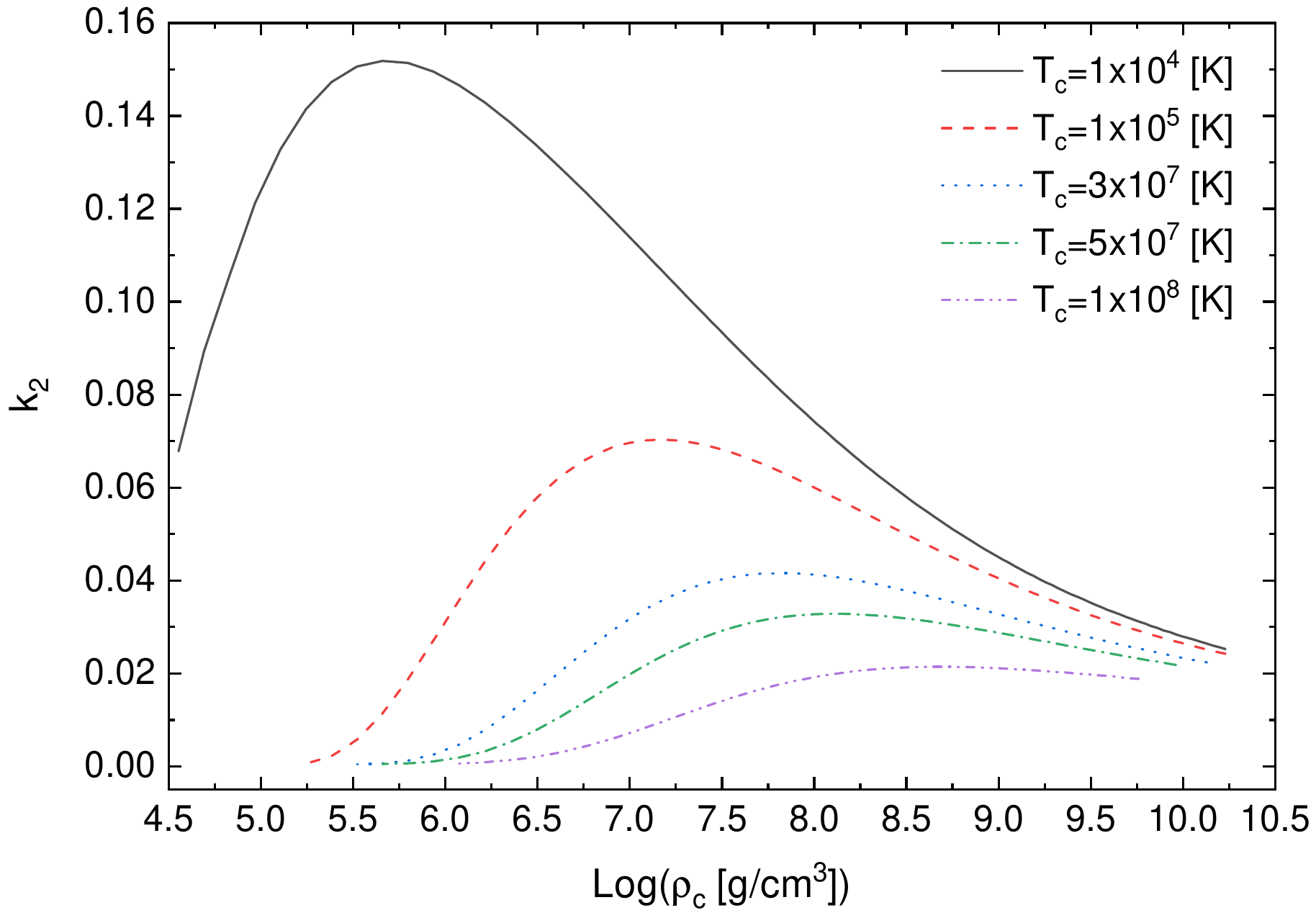}\\
\includegraphics[width=1\linewidth]{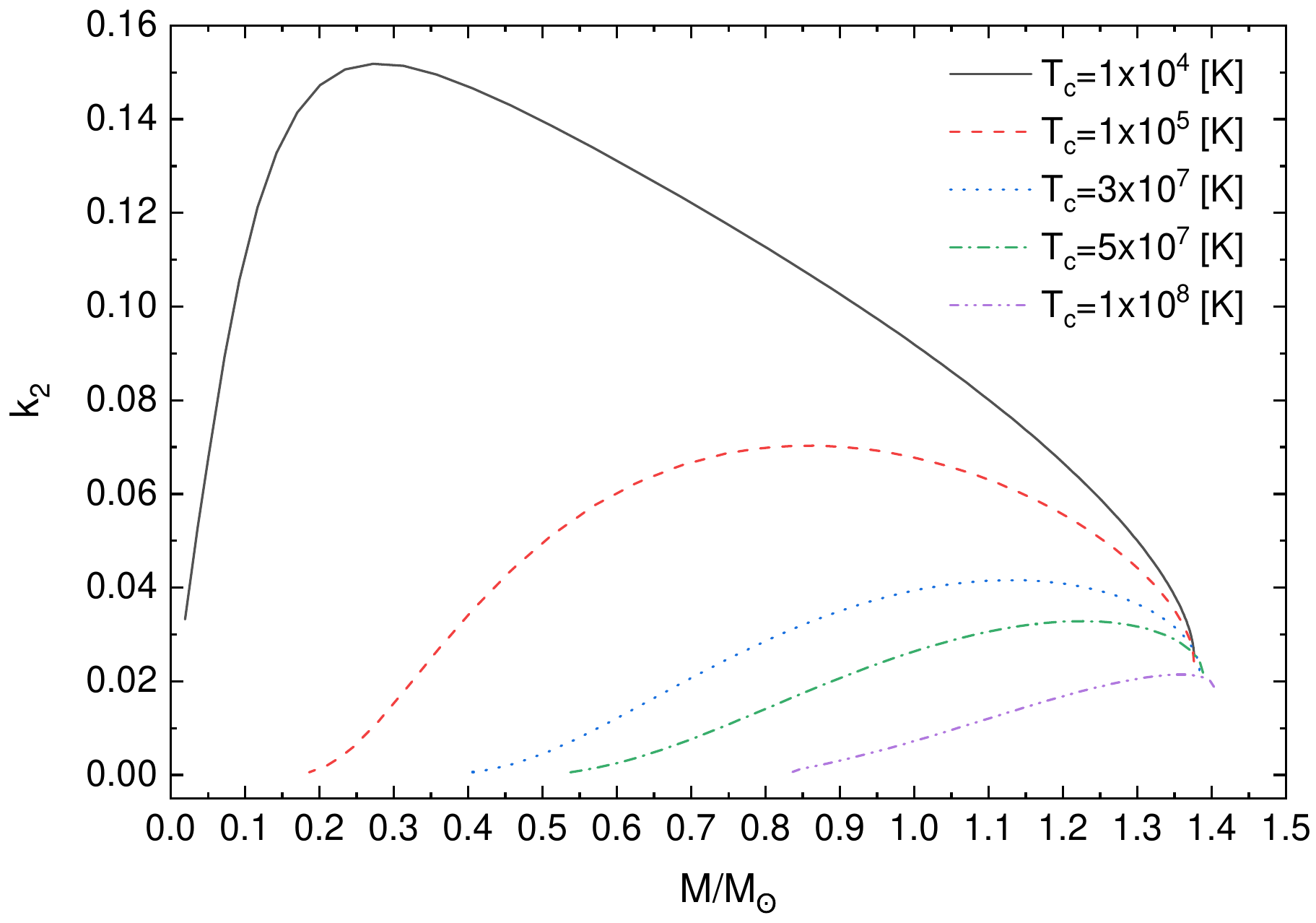}\\
\includegraphics[width=1\linewidth]{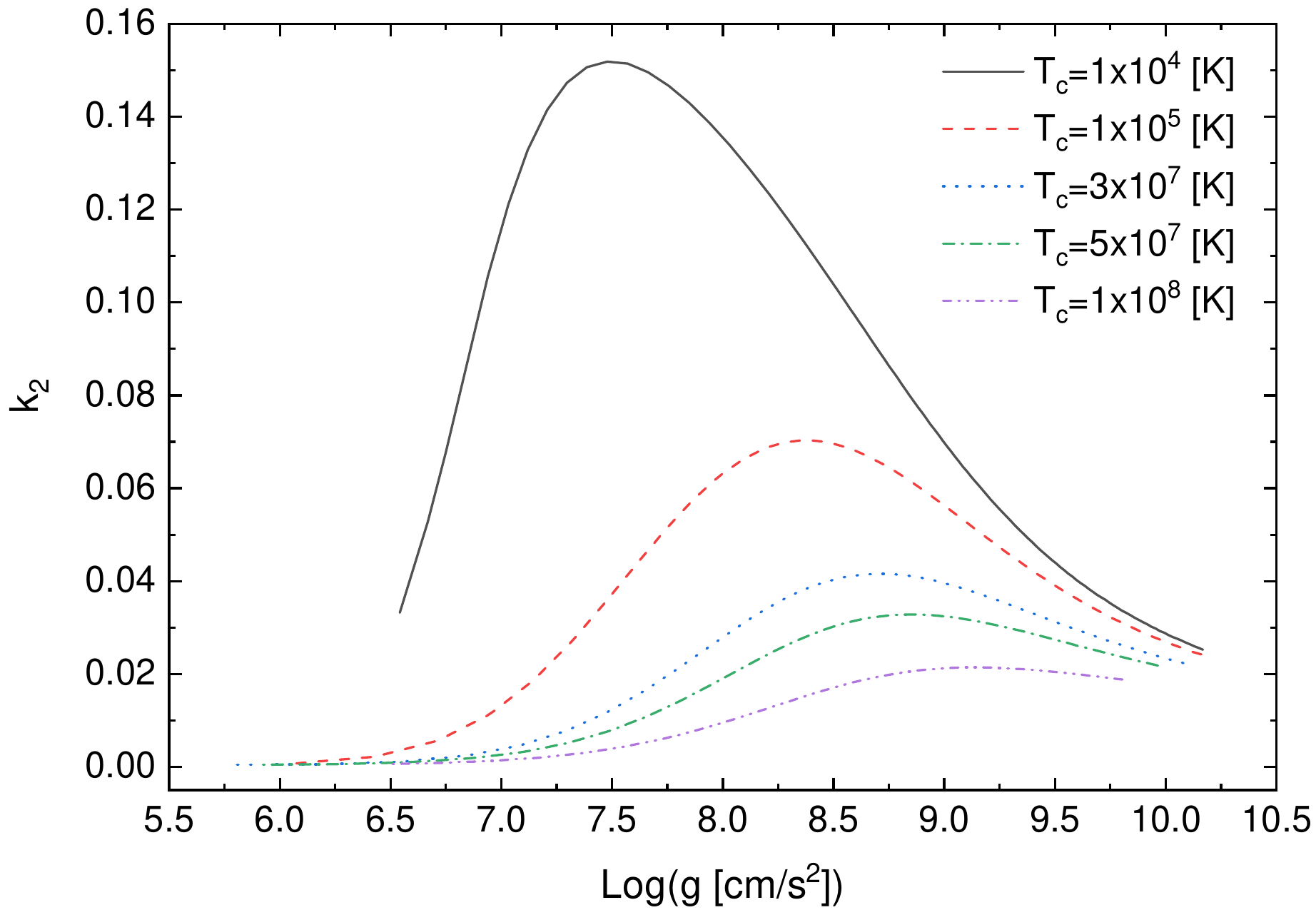}
\caption{The tidal Love number $k_2$ as a function of the central energy density $\rho_c$, total mass $M/M_{\odot}$, and surface gravity $g$ are respectively presented on the top, middle, and bottom panels for different central temperature values $T_c$. }
\label{figmxtcxk1}
\end{center}
\end{figure}

Thus, to investigate the influence of central temperature on the tidal Love number, we plot $k_2$ as a function of the central energy density $\rho_c$, the total mass $M/M_{\odot}$ (with $M_{\odot}$ representing the Sun's mass)  and surface gravity $g$ in Fig. \ref{figmxtcxk1} for different central temperature values.   { The selected range of central temperatures is based on the findings of Nunes et al. (2021). In this study, the authors estimated the central temperatures of several massive WDs, identifying values within the range $ \left(10^4 \, \text{K} \leq T_c \leq 10^8 \, \text{K}   \right)$ for observable massive stars. These values varied depending on the composition of the stellar envelope. Consequently, it is reasonable to extend the temperature range for analyzing \(k_2\) and \(\Lambda\) in the context of stellar structure. This broader range provides a robust and comprehensive framework for analysis.}  In all curves of the three panels, we note that the Love number grows with both central density, total mass, and surface gravity until it reaches a maximum value, where the curves turn clockwise so that $k_2$ starts to decrease with the increase of $\rho_c$, $M/M_{\odot}$, and $g$.  Moreover, in the panels, we also note the influence of $T_c$ on $k_2$. From some interval of central density, mass, and surface gravity we find that the Love number value decreases with the increase of the central temperature. This could be associated with the central pressure increasing with the central temperature, acting as an effective pressure to support more mass against gravitational collapse. Consequently, we have a more compact star.

\begin{figure}[!ht] 
\begin{center}
\includegraphics[width=1\linewidth]{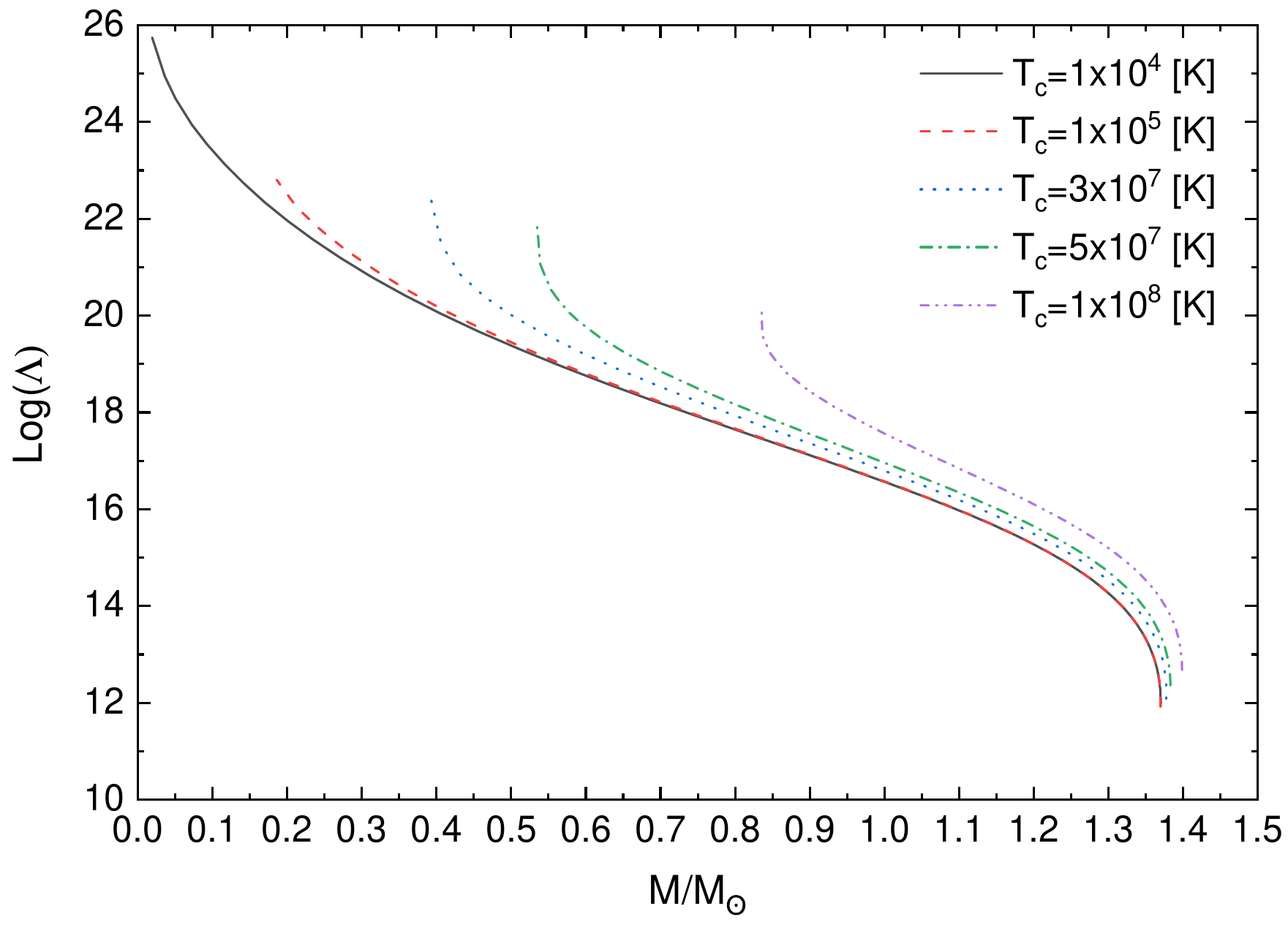}
\caption{Dimensionless tidal deformability against the total mass for some central temperature values. }
\label{figlamb_mass}
\end{center}
\end{figure}

The dimensionless tidal deformability $\Lambda$ versus the total mass is plotted in Fig. \ref{figlamb_mass} for different values of central temperatures. In all $T_c$ considered, we note that $\Lambda$ decays with increasing total mass. From this result, together with Fig. $4$ of the article \citep{Nunes_2021}, we can understand that the star has a lower distortion under the action of tidal forces as the radius decreases. This observation can play an essential role in understanding the thermodynamics of dense stellar objects, particularly in binary systems where tidal forces are significant. However, we note that dimensionless tidal deformability increases with central temperature in some range of total mass. This suggests that higher temperatures lead to a softer stellar structure, i.e., a stellar configuration that is more susceptible to deformation under external tidal forces. This relationship highlights how thermal properties can affect the internal stiffness of the star, influencing its response to tidal interactions. Since WDs begin their lives with relatively high core temperatures, the observation that higher temperatures correspond to greater tidal deformability supports the idea that as these stars cool over time, their structure becomes increasingly rigid; thus, having a configuration less prone to deformation. In other words, as the WD loses thermal energy and the internal pressure decreases, the star's response to tidal forces decreases. This thermal evolution plays an important role in how the star's physical properties, such as stiffness and deformability, change over its lifetime.

 { Beyond its role in stellar evolution, $\Lambda$ may significantly influence the detection of GWs. As shown in Figure \ref{figlamb_mass}, $\Lambda$ takes on smaller values for very massive WDs near the Chandrasekhar mass limit of $1.4 \, M_\odot$. This indicates that massive WDs are less prone to deformation, possessing a stiffer structure. Consequently, tidal interactions with a companion are more likely to produce GWs with amplitudes detectable by LISA.}

\subsection{Gravitational Waves and tidal deformability of hot WDs binary system}\label{Results1}

The chirp mass $M_c$ is an astrophysical parameter, particularly used when analyzing compact binary systems such as those composed of BHs, NSs, or WDs. This parameter represents a specific combination of the masses of the two objects in a binary system. It plays a crucial role in determining the frequency and amplitude of the GWs that the binary system emits. The chirp mass effectively combines information about the individual masses into a single value, simplifying the analysis of the gravitational waves emitted during the inspiral. In particular, the chirp mass influences the frequency and amplitude of the gravitational wave signal, allowing researchers to infer the mass properties of the binary system from the observed signals. At the lowest order of a post-Newtonian expansion, the phase evolution of the gravitational waveform is determined primarily by a specific combination of the masses of the two bodies, referred to as the chirp mass \citep{Cutler_1994}, denoted as
\begin{equation}\label{chirp}
M_c=\frac{(M_1M_2)^{3/5}}{(M_1+M_2)^{1/5}},
\end{equation}
where $M_1$ and $M_2$ correspond to the masses of each of the stars that make up the binary system.

\begin{figure}[!ht] 
\begin{center}\label{chirp_mass}
\includegraphics[width=1\linewidth]{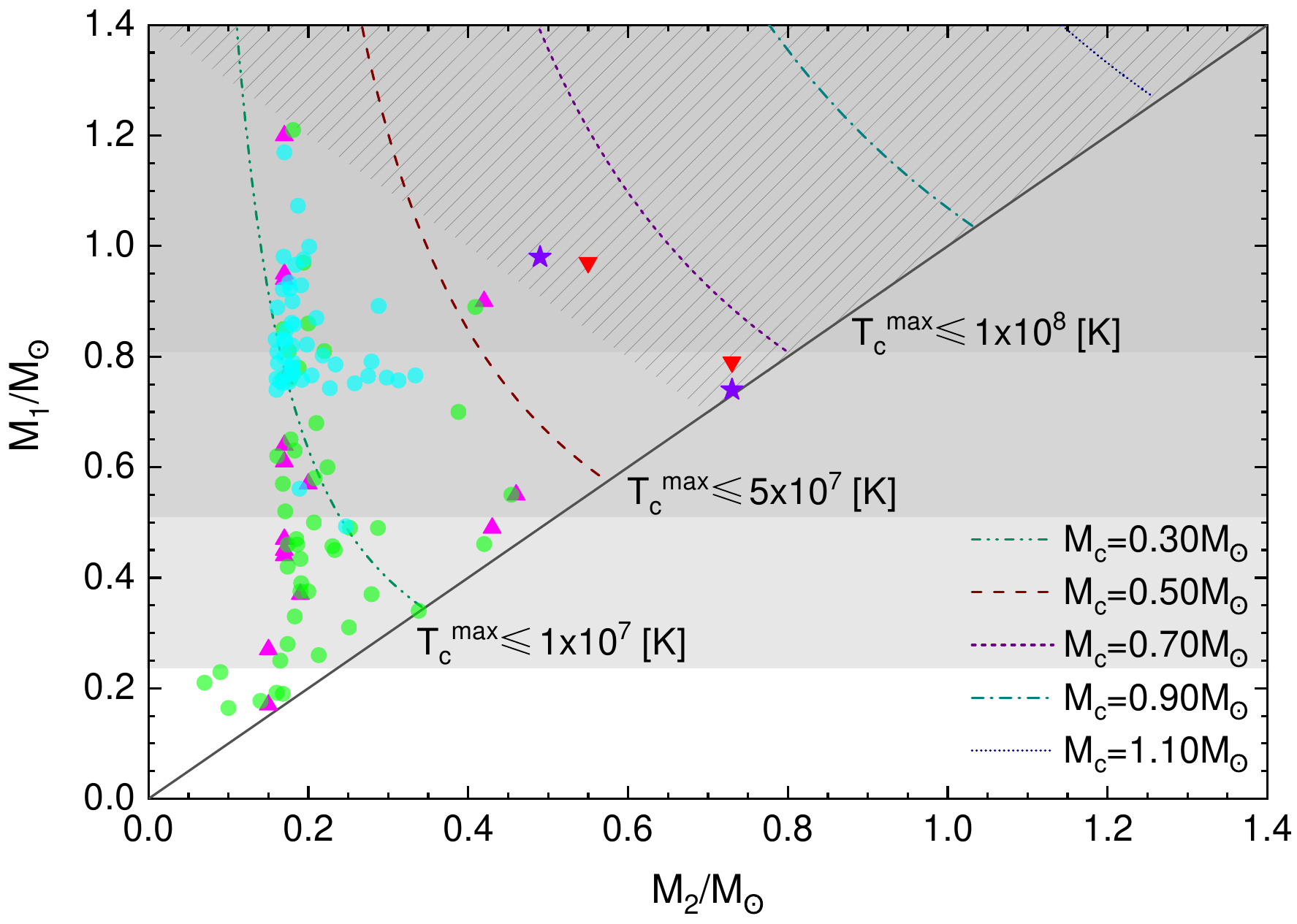}
\includegraphics[width=1\linewidth]{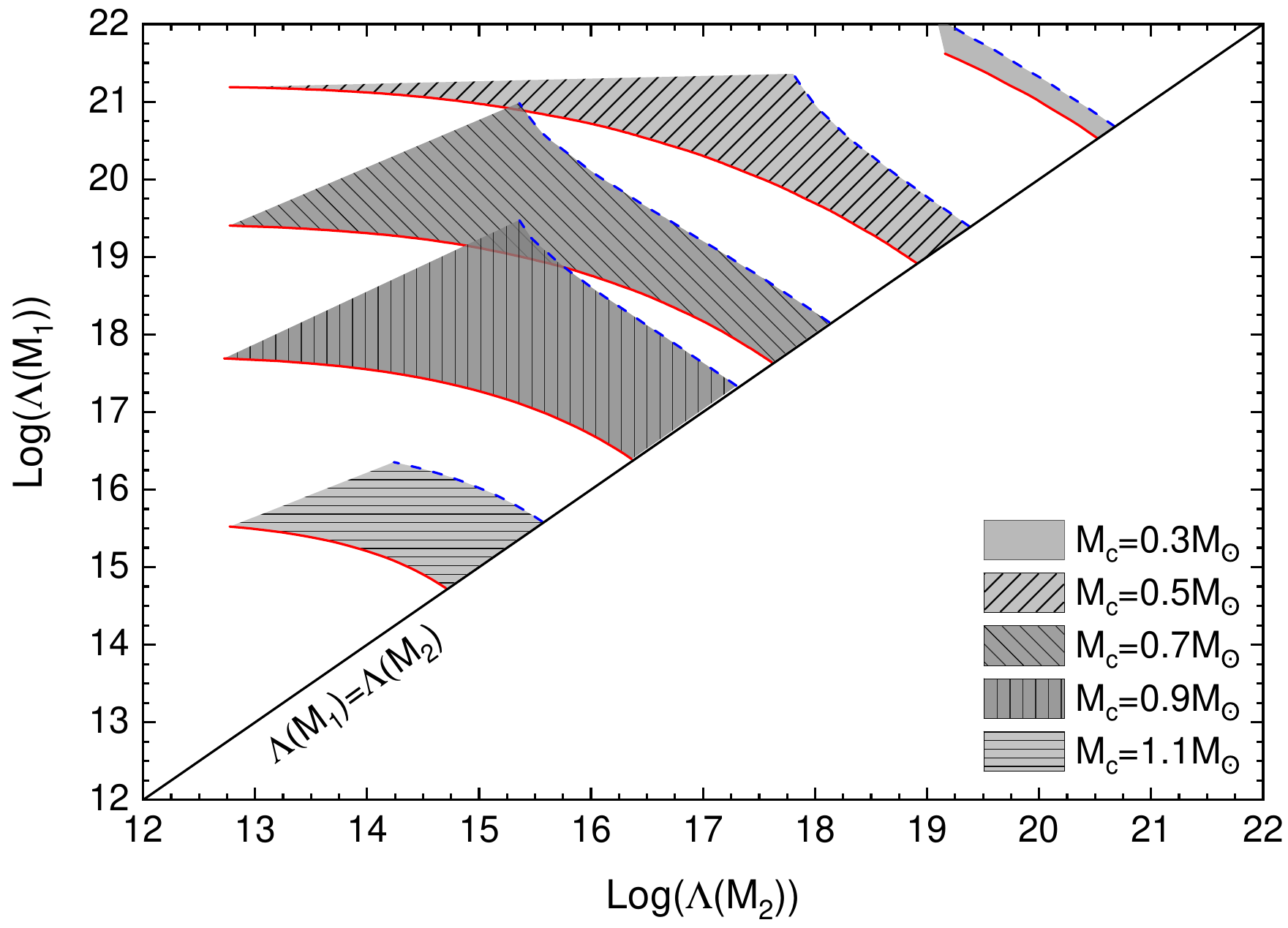}
\caption{Top panel: $M_1\times M_2$ for the different chirp masses and core temperatures of WDs. The solid diagonal line indicates the values corresponding to $M_1=M_2$. {The hashed region indicates binary masses above Chandrasekhar's mass limit}. Observational data extracted from the catalogs in \citep{Brown_2013}, \citep{Gianninas_2014}, and \citep{Brown_2016} are respectively marked with pink triangles, green circles, and cyan circles. The purple stars and the inverted red triangles correspond to the estimates of progenitor masses for J$2211+1136$ and J$1901+1458$~\citep{Sousa_2022}, respectively. Bottom panel: Dimensionless tidal deformability of two WDs in a binary system $\Lambda_1\times\Lambda_2$ with the diagonal line representing the values where $\Lambda_1 =\Lambda_2$. The red and dashed blue lines correspond to minimum and maximum central temperature values.}

\label{m1_m2}
\end{center}
\end{figure}
The diagrams $M_1\times M_2$ and $\Lambda_1\times\Lambda_2$ are plotted in the top and bottom panels of Fig.~\ref{m1_m2}, respectively, for different chirp masses and central temperatures. Based on the mass range and temperature effects described by \cite{Nunes_2021}, we selected chirp masses $M_c/M_\odot = \left[0.3, 0.5, 0.7, 0.9, 1.1\right]$ for our analysis. These values cover binaries ranging from low-mass to very massive WDs. These values are depicted by the green, wine, dark purple, dark cyan, and navy curves. The black diagonal line in the figure marks the point where $ M_1 = M_2$, corresponding to equal mass binaries.  In the top panel, the observational data of \cite{Brown_2013}, \cite{Gianninas_2014}, and \cite{Brown_2016} are represented by pink triangles, green and cyan circles, respectively. In addition, the estimations of J$2211+1136$ and J$1901+1458$ \citep{Sousa_2022} are marked as purple stars and inverted red triangles, respectively. The hashed region represents binaries with masses above the \cite{Chandrasekhar_1931} mass limit,  which corresponds to regions of instability \citep{Dan_2014}. Besides, these observations suggest that binaries are often found with at least one low-mass star. Since low-mass WDs have a larger radius and applying the Stefan-Boltzmann law, we deduce that they have higher luminosities than those of the more massive WDs at the same effective temperature. This higher luminosity makes low-mass binaries easier to detect, potentially explaining why massive binary WDs, which are dimmer, are not observed. On the bottom, $\Lambda_1\times\Lambda_2$ is plotted for both different chirp masses and different central temperature limits. Since different mass ranges support different minimum and maximum central temperature values \citep{Nunes_2021}, both the minimum and maximum central temperature values used can have different values. Thus, the solid red line indicates the minimum core temperature and the dashed blue line represents the maximum central temperature. The solid black diagonal line indicates where $\Lambda_1 = \Lambda_2$. The curves $\Lambda_1-\Lambda_2$, are constructed by choosing a value of $M_1$ and then finding $M_2$ through the chirp mass, by the equation (\ref{chirp}). The mass intervals for $M_1$ and $M_2$ change for each chirp mass value. For $M_c=0.3\,M_{\odot}$ we use $0.35\leq M_1/\,M_{\odot}\leq1.40$ and $0.11\leq M_2/\,M_{\odot}\leq0.34$, for $M_c=0.5\,M_{\odot}$ we employ $0.58\leq M_1/\,M_{\odot}\leq1.40$ and $0.25\leq M_2/\,M_{\odot}\leq0.57$, for $M_c=0.7\,M_{\odot}$ we consider $0.81\leq M_1/\,M_{\odot}\leq1.40$ and $0.49\leq M_2/\,M_{\odot}\leq0.80$, for $M_c=0.9\,M_{\odot}$ we have $1.04\leq M_1/\,M_{\odot}\leq1.40$ and $0.73\leq M_2/\,M_{\odot}\leq1.03$, and for $M_c=1.1\,M_{\odot}$ we take into account $1.27\leq M_1/\,M_{\odot}\leq1.40$ and $1.07\leq M_2/\,M_{\odot}\leq1.26$. From the figure, we also note the effects of the central temperature on the dimensionless tidal deformabilities $\Lambda_1$ and $\Lambda_2$. A higher core temperature leads to a larger tidal deformability. It follows from this that the tidal forces of a star can more easily distort its companion star in the binary system as the central temperature increases.

\begin{figure}[!ht] 
\begin{center}
\includegraphics[width=1\linewidth]{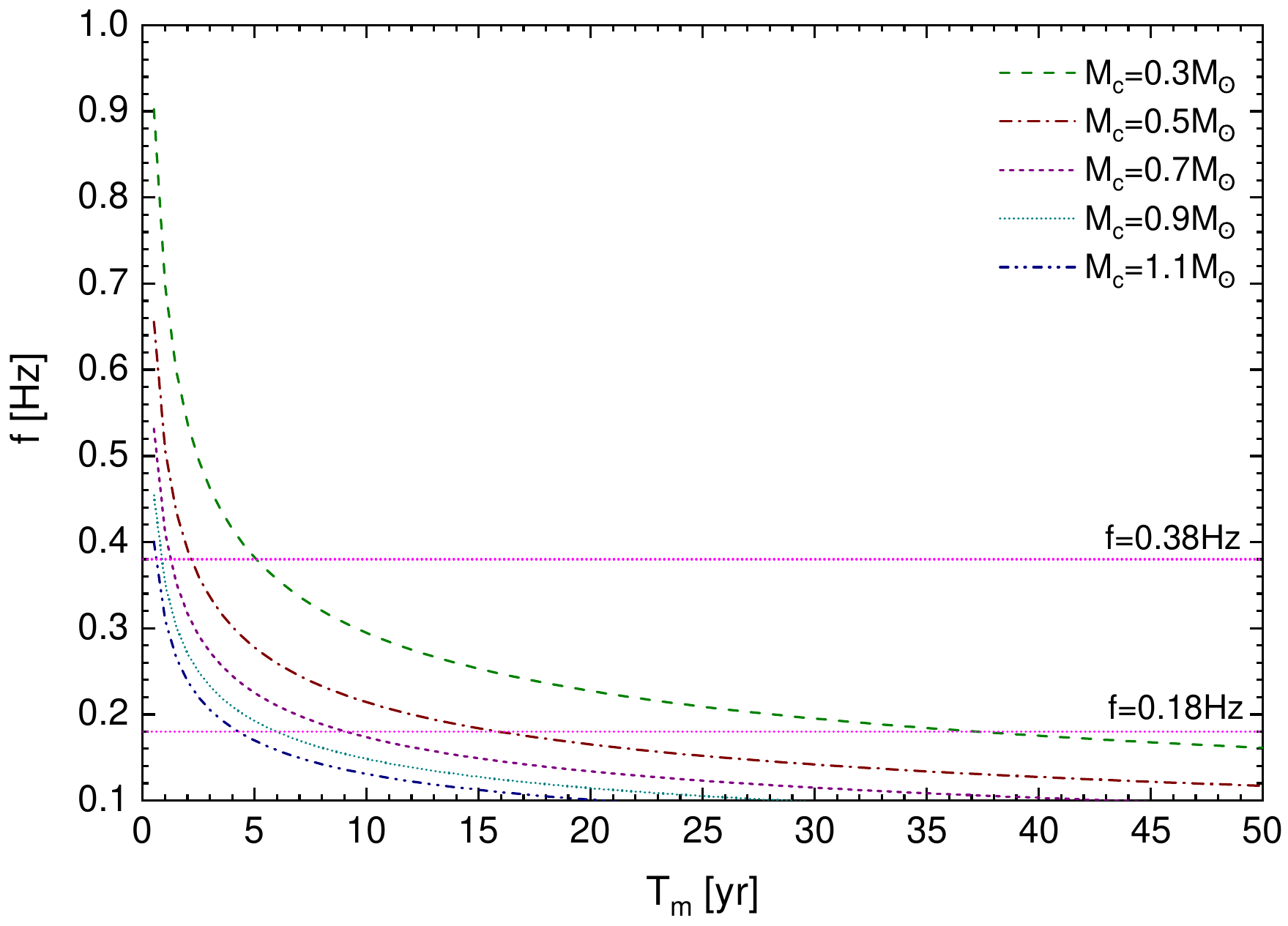}
\caption{Frequency as a function of evolution time for a WD binary system with several chirps masses. The dot pink line represents the critical frequency value for Gravitational wave detection \citep{Huang_2020}.}
\label{fxT}
\end{center}
\end{figure}

 { The evolution of the GW frequency (\(f\)) over time (\(t\)) is a crucial aspect in understanding the inspiral phase of the WD system. During this phase, the GW signal becomes progressively stronger and higher in frequency, becoming important for detection and analysis. For systems with circularized orbits, the relationship between frequency and time can be derived directly from the quadrupole approximation \citep{Maggiore2008}.} From equation \eqref{chirp}, the gravitational wave frequency $f$ evolves in time $t$ as \citep{Cutler_1994}
\begin{equation}
    \frac{d{f}}{dt}=\frac{96}{5}\pi^{8/3}M_c^{5/3}f^{11/3},
\end{equation}
which integrated gives us \citep{Huang_2020}
\begin{equation}\label{Eq_f}
    {f}=0.18\left(\frac{T_m}{\rm 5yr}\right)^{-3/8}\left(\frac{M_c}{1 M_\odot}\right)^{-5/8} \rm Hz,
\end{equation}
where $T_m$  { is the time remaining until the merger, with $T_m=0$ corresponding to the time of the merger itself.} The profile of the GW frequency with respect to $T_m$ is plotted in Fig.~\ref{fxT} for different values of chirp mass, namely, $M_c/M_\odot = \left[0.3, 0.5, 0.7, 0.9, 1.1\right]$. The figure shows a clear  { power-law decay, where the frequency decreases as $T_m$ increases. The} frequency range corresponds to LISA's sensitivity band, from $0.1$ mHz to $1$ Hz. Moreover, equation \ref{Eq_f}, as pointed out by \cite{Huang_2020}, defines a limit where the emitted GW can be considered monochromatic, meaning its frequency remains relatively stable over time, which enhances detectability. 

 { Beyond the monochromatic phase of the system's coalescence, there is another critical region to evaluate near \(T_m = 0\). This region corresponds to where the system reaches its maximum frequency just before the merger. We observe that this region is reached faster for systems with higher chirp masses than for those with lower chirp masses. }

 {This behavior arises since systems with higher chirp masses radiate gravitational waves more efficiently, leading to a more rapid loss of orbital energy and angular momentum. Consequently, the inspiral phase accelerates as \(M_c\) increases, bringing the system to higher frequencies within a shorter time frame. Besides, detecting the high-frequency signals in this region allows LISA to capture the final stages of the inspiral and even the pre-merger dynamics.  Moreover, the timescale of this region is within four years, aligning with LISA's planned initial operational timeframe.}

In addition, the amplitude of the gravitational wave is also crucial in determining whether the signal can be detected. Thus, to understand the relationship between gravitational wave amplitude and binary stars, we employ the definition \citep{Huang_2020} 
\begin{equation}\label{Eq_A}
    \mathcal{A}=\frac{2M_c}{d}\left(\pi f\right)^{2/3},
\end{equation}
where $d$ is the luminosity distance. We can observe that the chirp mass is proportional to the amplitude signal of the gravitational wave, suggesting that the larger the chirp mass, the larger the amplitude and the more detectable a source will be.

\begin{figure}[!ht] 
\begin{center}\label{lambda}
\includegraphics[width=1\linewidth]{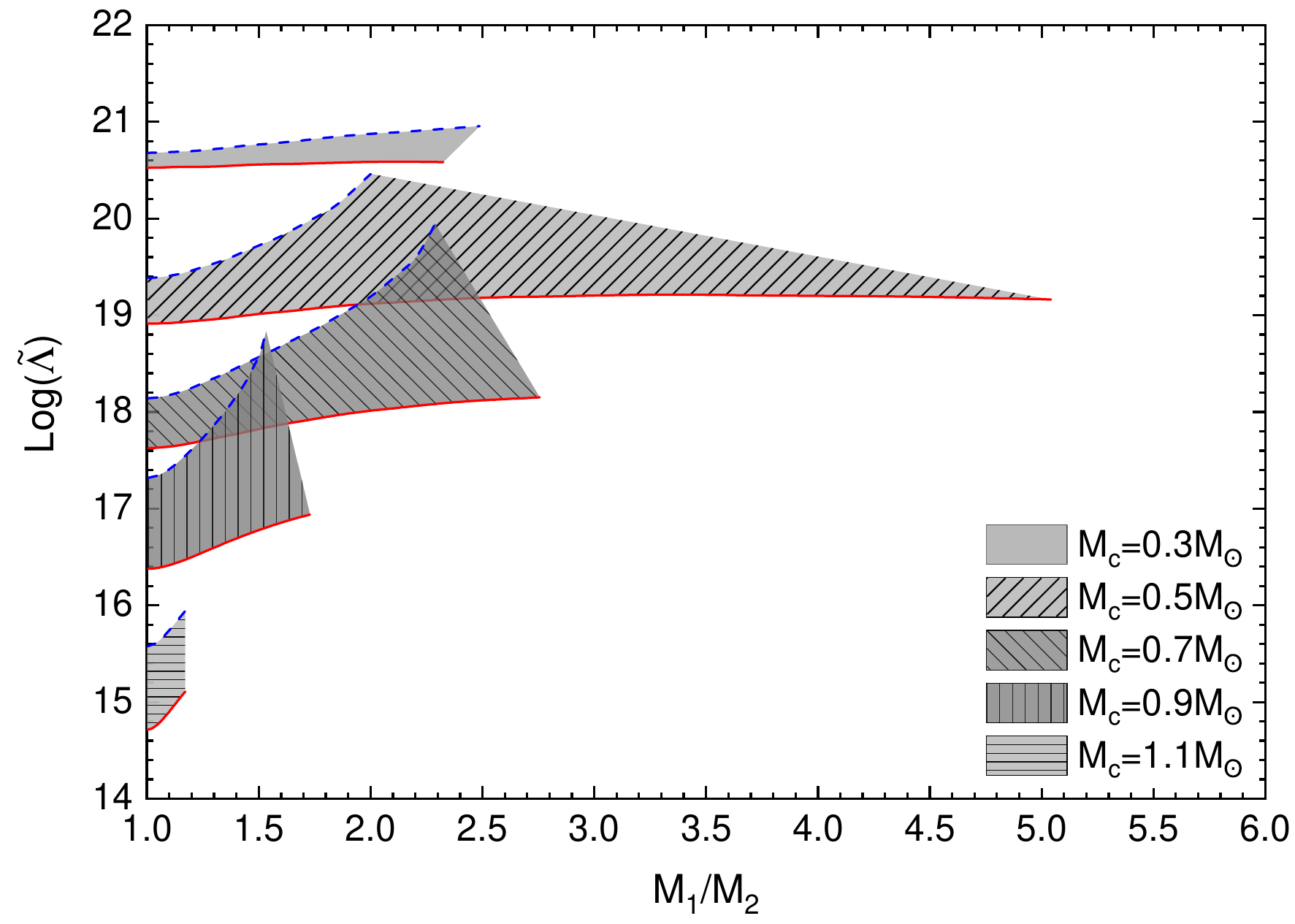}
\caption{The weighted tidal deformability ${\tilde\Lambda}$ against the ratio $M_1/M_2$. In both panels, different values of WD chirp masses and a range of central temperatures are considered.}
\label{lambda_q}
\end{center}
\end{figure}

In Fig. \ref{lambda_q} is plotted ${\tilde\Lambda}\times M_1/M_2$, for different chirp masses and two values of central temperature. In the figure, the dependence of the weighted dimensionless tidal deformability parameter ${\tilde\Lambda}$, which can be determined via gravitational waves in a binary system, is found as follows \citep{hinderer2008}:
\begin{equation}
    \tilde{\Lambda}
=\frac{16}{13}\frac{(M_1+12M_2)M_1^4\Lambda_1+(12M_1+M_2)M_2^4\Lambda_2}{(M_1+M_2)^5}.
\end{equation}

The figure shows that the weighted tidal deformability grows with temperature and decreases with the increment of the chirp mass. This behavior suggests that more massive systems may retain greater structural integrity under varying thermal conditions.

\section{Conclusion}\label{conclusion_ofthework}

With advancements in technology enabling the construction and upcoming launch of LISA, we anticipate gaining new insights and refining our understanding of the cosmos. This study explores conditions within the WD binaries as potential sources of gravitational waves, providing a basis for future observations. These findings aim to improve our predictive capabilities and contribute to the growing body of knowledge that will be tested once LISA begins its mission.

This is how we examined the central temperature effects on the tidal Love number and dimensionless tidal deformability of isolated WDs, as well as the frequency and amplitude of gravitational waves of WDs in a binary system. Our findings indicate that, for some range of total masses, the temperature effects result in lower tidal Love numbers and larger dimensionless tidal deformability values as the temperature rises. The change of the tidal Love number is more pronounced in low-density, low-mass stars, particularly those with surface gravity values  $\lesssim8\,\text{cm/s}^2$. In turn, the change of the dimensionless tidal deformability exhibits an increase with the central temperature. This suggests that higher central temperatures generate a softer internal structure, allowing the star to be more easily deformed by its companion.

Moreover, we contrasted the results obtained from  the binary hot WD system with the observational data extracted from the catalogs in \citep{Brown_2013, Gianninas_2014, Brown_2016}. We note that the cluster placed on low chirp masses indicates the presence of low-mass stars with large central temperatures. This prevalence of low-mass stars may suggest a correlation with stellar luminosity; according to the Stefan-Boltzmann law, lower-mass stars (with higher radius according to \cite{Nunes_2021}) tend to exhibit higher luminosity, which could enhance their detectability. Thus, it is essential to explore alternative detection methods for binaries with higher-mass stars, such as gravitational waves. In particular, some cases exhibit medium chirp mass values,  as indicated by the estimates of DWD progenitor masses for J$2211+1136$ and J$1901+1458$ \citep[see][]{Sousa_2022}. In addition, our results reveal that deformability is greater for smaller chirp masses than for more massive stars, thus suggesting that the gravitational wave frequency of more massive stars could be easier to detect due to their rigidity.

We also investigated binaries of WDs as potential sources of gravitational waves. Our analysis revealed that high-chirp mass binaries are more likely to be detected during the four-year operational period of LISA. { This behavior arises because systems with higher chirp masses radiate GWs more efficiently and produce waves with higher amplitudes.} When examining the weighted dimensionless tidal deformability parameter of WDs, we observed that these massive binaries maintain greater structural integrity. This finding aligns with previous studies of more compact stars \citep{2023PhRvD.107l4016A,2023EPJC...83..211A,2023JCAP...03..028L,2020arXiv200914274P} highlighting the importance of tidal deformability in understanding the detectability of gravitational waves from binary systems.

Finally, we conclude that future observations in the low-frequency range of LISA are crucial for investigating interactions among compact stars, particularly WDs.  By focusing on binary systems with high chirp masses, we expect to uncover new insights into the processes leading to Type Ia Supernovae, as these massive WDs play a key role in such events \citep{Filippenjo_1997, Wang_2012}.

\section*{Acknowledgments}
\noindent  { We thank the anonymous referee for their valuable suggestions and comments, which have helped improve this work.} JDVA would like to thank Universidad Privada del Norte and Universidad Nacional Mayor de San Marcos for the financial support - RR Nº$\,005753$-$2021$-R$/$UNMSM under the project number B$21131781$. Finally, C.H.L. would like to thank CNPq for the financial support with Projects No 401565/2023-8 (Universal CNPq) and No 305327/2023-2 (productivity program).  J.G.C. is grateful for the support of FAPES (1020/2022, 1081/2022, 976/2022, 332/2023), CNPq (311758/2021-5), and FAPESP (2021/01089-1).

\bibliographystyle{apalike}  

\bibliography{bibliography.bib}

\end{document}